\documentclass[12pt]{article}

\usepackage{scicite}
\usepackage{caption}
\usepackage{times}
\usepackage{afterpage}
\usepackage{floatpag}
\usepackage{graphicx} 

\newenvironment{sciabstract}{%
\begin{quote} \bf}
{\end{quote}}

\title{Attosecond metrology in circular polarization}

\author
{Meng Han,$^{1\ast}$ Jia-bao Ji,$^{1}$ Kiyoshi Ueda, $^{1,2}$ Hans Jakob Wörner$^{1}$\\
\\
\normalsize{$^{1}$Laboratorium für Physikalische Chemie, ETH Zürich, Zürich, 8093, Switzerland}\\
\normalsize{$^{2}$Department of Chemistry, Tohoku University, Sendai, 980-8578, Japan}\\
\normalsize{$^\ast$To whom correspondence should be addressed; E-mail: meng.han@phys.chem.ethz.ch}\\
}

\date{}

\begin{document} 


\baselineskip24pt


\maketitle


\begin{sciabstract}
Attosecond metrology with linearly polarized light pulses is the basis of a highly successful research area. An even broader impact can be expected from a generalized metrology that covers two-dimensional polarization states, enabling notably the study of chiroptical phenomena on the electronic time scale. Here, we introduce and demonstrate a comprehensive approach to the generation and complete characterization of elliptically to circularly polarized attosecond pulses. The generation relies on a simple plug-in device of unprecedented simplicity. For the characterization, we introduce SPARROW (Stokes-Parameter and Attosecond Resolved Reconstruction of Optical Waveforms), which encodes the attosecond-metrology information into the photoemission angle in the polarization plane and accesses all four Stokes parameters of the attosecond pulses. Our study demonstrates a physically transparent scheme for attosecond metrology with elliptical to fully circular polarizations, applicable to both table-top and accelerator-based light sources, which will unlock studies of chiral molecules, magnetic materials and novel chiroptical phenomena on the most fundamental time scales. 
\end{sciabstract}

\clearpage
High-harmonic generation (HHG) from noble gases driven by intense femtosecond lasers offers bright coherent extreme-ultraviolet (XUV) light sources, which deliver attosecond pulse trains or isolated attosecond pulses. These attosecond light sources have advanced ultrafast science into the attosecond time domain ranging from atomic \cite{Hentschel2001,Paul2001} and molecular \cite{sansone10a,Huppert2016} systems, to solids \cite{Ghimire2011,Hammond2017,Lu2019,uzan20a} and liquids \cite{Luu2018,Jordan2020}. One longstanding challenge about HHG light sources is the polarization control of the attosecond light pulses. The recollision mechanism of the HHG process precludes the generation of circularly polarized attosecond light pulses by simply using circularly polarized driving fields. On the other hand, the polarization conversion \cite{hochst1994} after generation relies on the polarization sensitivity of metal surfaces, limiting the wavelength range and bandwidth of the circularly polarized radiation. With the scope of attosecond science extending to chiral molecules \cite{Beaulieu2017,baykusheva18a}, ultrafast magnetism \cite{bandrauk2017circularly,yuan2013single,yuan2020ultrafast,Siegrist2019} and topological materials \cite{schmid21a,baykusheva21a}, tabletop light sources for circularly polarized attosecond pulses can be expected to drive major progress.

At large-scale facilities, such as femtosliced synchrotrons \cite{Bahrdt1992,N2011} and free-electron lasers (FELs) \cite{Allaria2012}, circularly polarized XUV pluses with pico- to femtosecond durations have been accessible for some time. Circularly polarized attosecond pulses have so far only been within reach of HHG sources. For example, using collinear two-color ($\omega$ + $2\omega$) counter-rotating circularly polarized ("bicircular") driving fields, the produced harmonics are circularly polarized but have opposite helicities for two adjacent orders \cite{Long1995,Eichmann1995,Fleischer2014,Kfir2014,kfir2016line}. In the time domain this corresponds to a pulse train of three linearly polarized attosecond bursts per laser cycle, where the polarization rotates $120^\circ$ between each burst \cite{Chen2016,Carlos2016a,baykusheva16a}. Resonant HHG of an elliptical laser field can produce bright low-order harmonics with an ellipticity of up to 0.7 \cite{Ferre2015} or 0.86 \cite{svoboda2022generation}. The coherent superposition of two orthogonal linearly-polarized HHG beams was demonstrated to yield circular polarizations \cite{Ellis2018,Azoury2019}. 
In a recent breakthrough, Hickstein et al. \cite{Hickstein2015} showed that two counter-rotating driving fields in a non-collinear geometry can produce two circular XUV beams with different helicities. When the two driving pulses were compressed to few optical cycles, the produced HHG spectrum was demonstrated to be continuous \cite{Huang2018} and the corresponding circularly polarized attosecond pulse is expected to be isolated. However, the implementation of the non-collinear HHG scheme is complex, and the temporal characterization is missing. Different from the case of linearly polarized attosecond pulses, the temporal characterization of an attosecond pulse with a two-dimensional (2D) polarization state requires a measurement of the phase difference between its two orthogonal polarization components. Therefore, the complete characterization of a circularly polarized sub-femtosecond pulses, i.e. including their temporal structure \cite{Chen2016,jimenez2018attosecond} and polarization state \cite{barreau2018evidence,svoboda2022generation,mazza2014determining,kazansky11a}, has remained out of reach so far, due to the lack of suitable metrologies.

\begin{figure}[htbp]
\centering
\includegraphics[width=\linewidth]{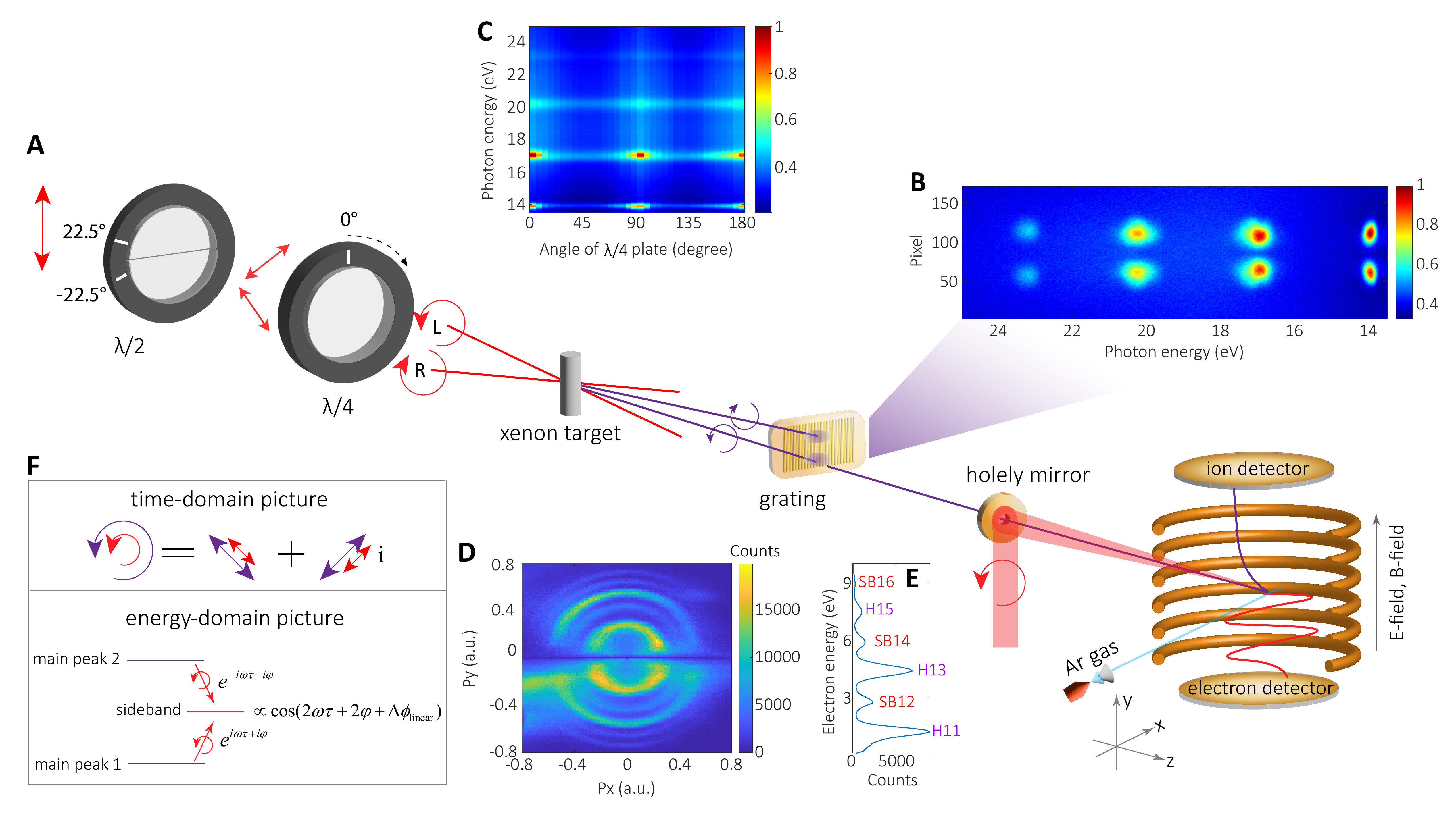}
\caption{\textbf{Generation of circularly polarized attosecond light pulses}. \textbf{A}, Experimental setup. The half-waveplate is custom designed to consist of two half disks whose fast axes form an angle of 45$^{\circ}$, as marked by the white lines on the holder. After passing through this half-waveplate, one beam will be divided into two beams with orthogonal polarizations. The following quarter-waveplate will convert the polarizations of the two beams to be circularly polarized with opposite helicities. \textbf{B}, Typically measured XUV spectrum generated from xenon gas. The two XUV beams have almost equal intensity but opposite helicities. \textbf{C}, Measured XUV spectrum as a function of the fast-axis angle of the quarter-waveplate. Rotating the quarter-waveplate by 90 degrees will exchange the helicities of the two IR beams as well as those of the two XUV beams. Note that the maximum XUV intensity is achieved at the circular driving fields, which is in contrast to the conventional HHG experiments driven by one IR beam. \textbf{D}, Measured delay-integrated photoelectron momentum distribution in the polarization plane by COLTRIMS. The discontinuity at Py = 0 is from the intrinsic limitation of the COLTRIMS spectrometer. \textbf{E}, Measured delay-integrated photoelectron energy spectrum along the emitting direction of $\mathrm{arctan}(Px/Py) < 5^\circ$. \textbf{F}, Time-domain and energy-domain pictures of characterizing the circularly polarized attosecond light pulse using the circular RABBIT experiment with co-rotating XUV and IR fields. In the time domain, the circular RABBIT experiment is equivalent to two linear RABBIT experiments performed along the orthogonal directions, and each linear RABBIT experiment can be used to characterize the XUV-field component along its direction. In the energy domain, the sideband is formed by the interference between the electronic states with different magnetic quantum numbers, so that the azimuthal angle $\varphi=\mathrm{arctan}(Px/Py)$ plays the same role as the XUV-IR time delay $\tau$, manifesting the principle of attosecond angular streaking.}
\label{fig:figure1}
\end{figure}

Here, we introduce a general approach to circular attosecond metrology, which overcomes all of the above-mentioned challenges. We generate attosecond pulses with tunable ellipticities ranging up to circularity based on a plug-in apparatus of unprecedented simplicity. We demonstrate their complete characterization in terms of their attosecond temporal structure and all 4 Stokes parameters. As illustrated in Fig. \ref{fig:figure1}A, our apparatus consists of a custom-designed half-waveplate and a conventional quarter-waveplate. The half-waveplate is composed of two half disks, whereby the relative angle between the fast axes of the two parts is $45^\circ$. After passing through this half-waveplate, the polarization directions of the upper and lower parts of the beam are orthogonal to each other. The quarter-waveplate then converts them to elliptical to circular polarizations with opposite helicities. The idea of our apparatus is to create the subbeams in one beam for the realization of the non-collinear geometry, and thus we name it beam-in-beam apparatus. When focusing this driving beam onto a gas target (xenon), the two generated XUV beams are almost identical in terms of profile, flux and spectrum, as illustrated in the measured XUV spectrum shown in Fig. \ref{fig:figure1}B. When the two counter-rotating driving beams are both circularly polarized with opposite helicities, their coherent superposition on the target has linear polarization with a direction that rotates across the focal spot, such that the generated XUV beams will be purely circularly polarized with the highest flux. Our beam-in-beam apparatus naturally satisfies this requirement due to the perfect achievable symmetry of the driving beams. Furthermore, there is no relative temporal and spatial jitter between the two driving IR pulses.

Figure \ref{fig:figure1}C illustrates the measured XUV spectrum as a function of the fast-axis angle of the quarter-waveplate. At the angle of $45^\circ$ or $135^\circ$, the two driving beams are linearly polarized along the orthogonal directions such that their coherent superposition gives rise to the circular or elliptical polarization on the target and thus the HHG intensity is the lowest, even though the produced HHG field is linearly polarized in the far field. Note that here the HHG intensity reaches a maximum when its polarization is most circular, which is opposite from the traditional ellipticity-resolved HHG experiments where the HHG flux is maximal at linear polarization \cite{antoine1997polarization}. Using our beam-in-beam apparatus, switching the helicity of the XUV beam is very straightforward. Rotating the quarter-waveplate by $90^\circ$ will exchange the helicities of the two driving IR beams, as well as those of the two produced XUV beams. After picking up one XUV beam for further characterization or application, our apparatus allows to vary the XUV helicity by rotating the quarter-waveplate without altering the XUV beam path. Regarding the crossing angle between the two driving beams, it can be adjusted by varying the focal length of the used focusing element. Here the crossing angle is controlled at 25 mrad, and such a small angle can ensure that the higher-order XUV beams are largely suppressed \cite{bertrand11a}, such that the $\pm 1$-order beams dominate \cite{Huang2018}.

Using a perforated mirror to select one XUV beam and recombining it with a dressing IR field, we demonstrate a scheme to fully characterize circularly and elliptically polarized attosecond light pulses, including its polarization state and temporal structure that we call SPARROW (Stokes-Parameters and Attosecond Resolved Reconstruction of Optical Waveforms). The first part of SPARROW is to characterize the harmonic phases in a co-rotating circularly polarized IR field, which generalizes the concepts of RABBIT \cite{Paul2001} (Reconstruction of attosecond beating by interference of two-photon transitions) and angular streaking \cite{Eckle2008,Han2021}. For detection, we use a COLTRIMS (Cold target recoil ion momentum spectroscopy) spectrometer \cite{Ullrich2003,Dorner2000}, but a co-axial velocity-map-imaging spectrometer \cite{li18a} or other equivalent detectors \cite{allaria14a} could also be used. Previous RABBIT experiments were limited to linearly polarized fields. Here, we experimentally demonstrate circular RABBIT with argon serving as the sample gas. The measured photoelectron momentum distribution in the polarization plane is shown in Fig. \ref{fig:figure1}D, where the XUV-IR phase delay was integrated over. The corresponding energy spectrum is depicted in Fig. \ref{fig:figure1}E. Three main peaks (H11, H13 and H15) created by the XUV ionization and two sidebands (SB12 and SB14) by the two-photon (XUV and IR) interactions are clearly observed. Our detector has no electron resolution along the transverse momentum direction of Py. In spite of this, the angular distributions of main peaks and sidebands are clearly more isotropic in comparison with the distributions obtained in the linear XUV case. The photoelectron energy spectrum contained the information of the XUV spectral intensity in the frequency domain. Further temporal characterization in the attosecond time domain is still needed.

As illustrated in Fig. \ref{fig:figure1}F, the circular RABBIT measurement decouples the 2D polarization of the XUV field into two orthogonal components and measures the harmonic phase difference between two adjacent orders. In the linear RABBIT case, the sidebands can be interpreted by the interference between the two-photon transition amplitudes from the two adjacent main peaks to the middle sideband, and thus the sideband yield is expressed as $Y_\mathrm{SB} = |a_\mathrm{up}(\theta)e^{i\omega\tau} + a_\mathrm{down}(\theta)e^{-i\omega\tau}|^2 \propto A+B\mathrm{cos}(2\omega\tau+\Delta\phi_\mathrm{linear})$, where $\omega$ is the center frequency of the dressing IR field, $\tau$ is the XUV-IR delay, $\theta$ is the electron emission angle with respect to the co-polarization direction, and $\Delta\phi_\mathrm{linear}$ is the phase difference between the two transition amplitudes [$a_\mathrm{up}(\theta)$ and $a_\mathrm{down}(\theta)$]. $\Delta\phi_\mathrm{linear}$ will be referred to as RABBIT phase because it can be directly obtained by sine-fitting or Fourier transformation of the sideband yield along the delay axis. The RABBIT phase consists of two parts, i.e., the finite derivative of the XUV spectral phase $\Delta\phi_\mathrm{XUV}$ (reflecting the attochirp) and the atomic phase difference $\Delta\phi_\mathrm{atomic}$. The attochirp is used to characterize the attosecond temporal structure of the XUV field \cite{Paul2001,mairesse2003}, while the atomic phase is related to the dynamics of the XUV+IR two-photon photoionization process \cite{klunder2011}. Here, the circular RABBIT technique involves the interference between the electronic states with different magnetic quantum numbers $m$ since absorbing (emitting) one left-circular IR photon will increase (decrease) $m$ by 1, such that not only the phase of the transition matrix elements but also the angular phase ($e^{im\varphi}$) of the populated continuum states must be taken into account. Thus, the sideband yield becomes $Y_\mathrm{SB} = |a_\mathrm{up}(\theta)e^{i\omega\tau+i\varphi} + a_\mathrm{down}(\theta)e^{-i\omega\tau-i\varphi}|^2 \propto A+B\mathrm{cos}(2\omega\tau+2\varphi+\Delta\phi_\mathrm{linear})$, where $\varphi = \mathrm{arctan}(Px/Py)$ is the electron emission angle in the polarization plane and $\theta$ specifies the angle with respect to the light propagation direction. It thus becomes clear that the photoelectron emission angle $\varphi$ and the XUV-IR time delay $\tau$ play equivalent roles. Therefore, the RABBIT phase can be obtained by resolving the electron emission angle, instead of scanning the XUV-IR time delay.

\begin{figure}[htbp]
\centering
\includegraphics[width=\linewidth]{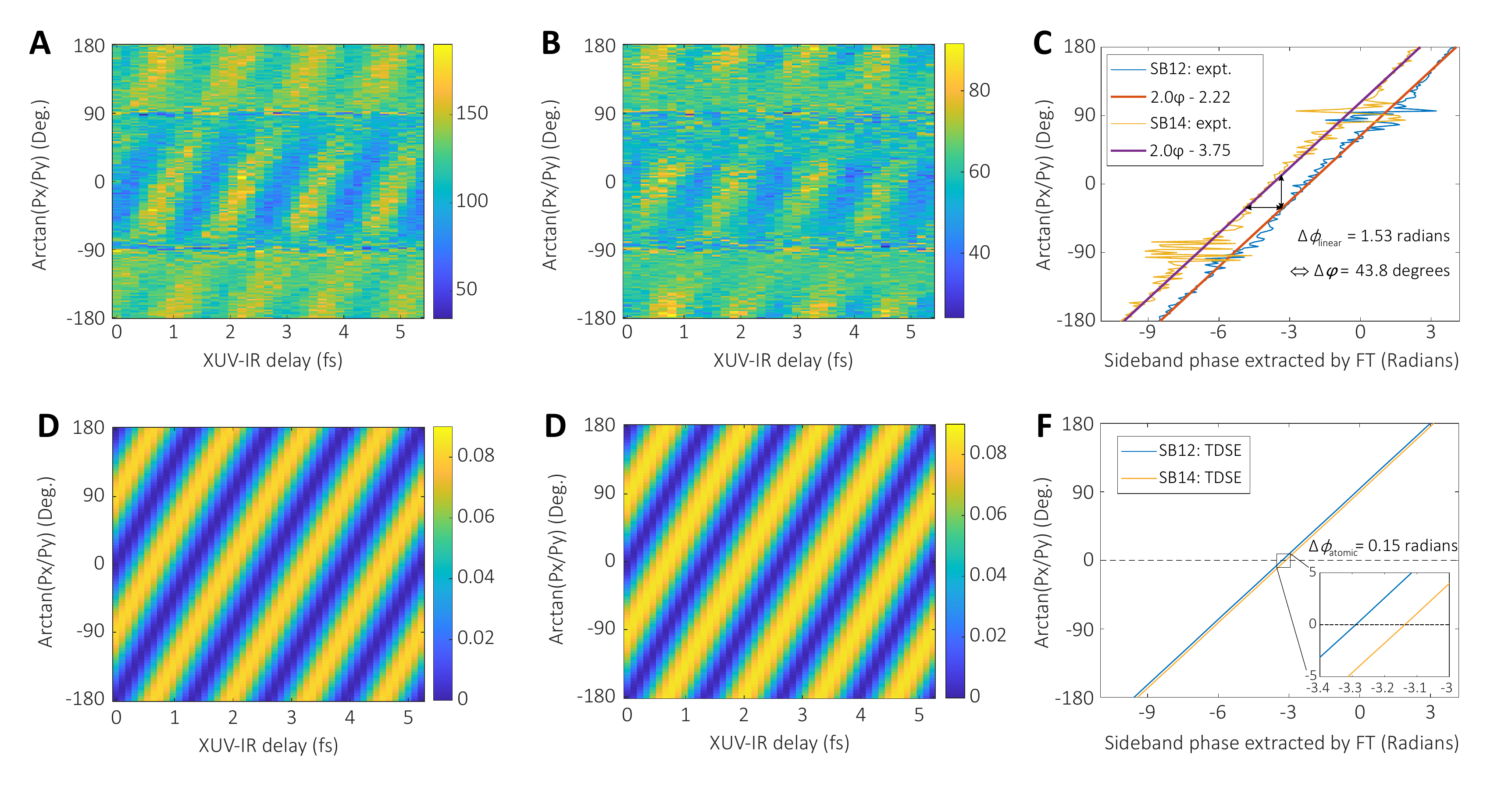}
\caption{\textbf{The equivalence of emission angle and phase delay in circular RABBIT, the first part of SPARROW}. Measured time-resolved angular distributions of SB12 (\textbf{A}), and SB14 (\textbf{B}). \textbf{C}, Angle-resolved sideband phase extracted by Fourier transformation along the axis of the XUV-IR delay. The phase of each sideband has a clear linearity with respect to the emission angle $\varphi = \mathrm{Arctan}(Px/Py)$. The phase delay between the two sidebands can be obtained from the extracted sideband phase difference at the same emission angle or from the emission angle difference at the same XUV-IR delay. \textbf{D-F}, Corresponding results from TDSE simulations averaged over p$_{+}$, p$_{0}$ and p$_{-}$ orbitals for the ground state. In the TDSE simulations, the XUV field is unchirped such that the tiny phase difference ($\sim$0.15~radians) between the two sidebands can be attributed to the energy dependence of the atomic phase in photoionization.}
\label{fig:figure2}
\end{figure}

To illustrate the angular-streaking principle in our circular RABBIT experiments, in Figs. \ref{fig:figure2}A-B we show the time- and angle-resolved photoelectron spectra for SB12 and SB14, where the electron kinetic energy is integrated within [2.2, 3.4] eV and [5.2, 6.6] eV, respectively. The slanted stripes in the spectra show that the sidebands are rotating in the co-polarization plane when varying the XUV-IR time delay. At each emission angle $\varphi$ the sideband yield oscillates along the axis of the time delay $\tau$ with a period of 1.33 fs, and remarkably this oscillation of the sideband yield has a linearly varying phase with respect to the angle. Performing a Fourier transformation along the time-delay axis, we obtained the angle-resolved phases of the oscillations of SB12 and SB14, respectively, as illustrated in Fig. \ref{fig:figure2}C. Fitting the phases with a linear function of $\varphi$, we determined the slope and intercept of each sideband phase. Both slopes are 2.0 within the 95$\%$ confidence interval, which validates our previous analysis of the sideband phase in circular RABBIT measurements, i.e., it confirms that the expression $2\varphi+\Delta\phi_\mathrm{linear}$, is correct. This also indicates that the attochirps of the two orthogonal polarization components should be equal. There are two equivalent methods to determine the difference between the RABBIT phases of SB12 and SB14. The ordinary method is to compare the two sideband phases at a fixed angle, i.e., comparing the two curves in Fig. \ref{fig:figure2}C horizontally. The other approach is to determine the emission angle difference at a fixed time delay, i.e., comparing the two curves vertically. Our experimental results reveal that the RABBIT phase difference between SB12 and SB14 is 1.53 radians or 43.8 degrees, which confirms the equivalence between the time delay $\tau$ and the emission angle $\varphi$. Thus, the circular RABBIT/angular streaking experiment allows to perform attosecond metrology without the need to scan the XUV-IR time delay, which is suitable for single-shot measurements. Such single-shot measurements cannot be realized in COLTRIMS using table-top sources, but they are highly relevant for attosecond metrology in circular polarization at FELs, where multi-hit-capable detectors are routinely used \cite{maroju20a,duris20a}.

In principle, here the observed RABBIT phase difference between SB12 and SB14 could be attributed to the atomic phase and/or the attochirp of the XUV field. To disentangle them, we resort to the \textit{ab-initio} solution of the time-dependent Schr\"odinger equation (TDSE) in the single-active-electron approximation. In the simulation, both the XUV and IR fields have no chirps and the simulated results shown in Fig. \ref{fig:figure2}D-F are averaged over p$_{+}$, p$_{0}$ and p$_{-}$ orbitals for the ground state. More details about the simulation are given in the Supplementary Materials (SM). From the time- and angle-resolved photoelectron spectra for SB12 and SB14, it is difficult to observe the existing difference. After Fourier transformation, the obtained phase difference between SB12 and SB14 can be accurately determined. It is very small, about 0.15 radians (see the enlarged plot in Fig. \ref{fig:figure2}F), which can only originate from the atomic phase. Therefore, in our pulse reconstruction process we subtract this small atomic phase shift from the measured RABBIT phase and then obtain the finite difference of the XUV spectral phase.

To further illustrate the circular RABBIT method, in SM (Fig. S1) we illustrate the RABBIT phase distribution in the whole momentum plane, which is obtained by Fourier transformation at each momentum with respect to the XUV-IR phase delay. The obtained phase spectrum is also called the phase-of-the-phase spectrum \cite{Skruszewicz2015,Han2021}, which can resolve the energy-dependent feature within one sideband (i.e., similar to the rainbow RABBIT \cite{Gruson2016}), and the offset angle in the manner of angular streaking. The energy dependence within one sideband is largely determined by the femtosecond structure of the XUV field \cite{Isinger2019}, which will be addressed in future studies.

When the XUV field is elliptically polarized, the polarization vector rotates in space with a non-uniform angular speed. Through single-photon ionization, this non-uniform rotation feature will be imprinted onto the photoelectron phase at the main peak and then onto the RABBIT phase at sidebands after interacting with a circular dressing IR field. In this case the RABBIT phase is not linearly varying with respect to the emission angle, and the discrepancy is directly related to the non-uniform rotation feature of the XUV polarization vector. In SM (Fig. S2), we illustrate the experimental results recorded at the ellipticity magnitude of 0.84 and 0.58 for the driving IR fields, which corresponds to that the fast-axis angle of the quarter-wave plate amounting to 5 and 15 degrees, respectively. The results show that the angle-resolved sideband phase will deviate from the linearity when decreasing the ellipticity. In SM, we also illustrate the general case where the two adjacent harmonics have different ellipticities or tilt angles (in other words, the attochirps are not equal for two orthogonal polarization components), in which the deviation of the angular phase from the linearity can directly manifest the sub-cycle property of the 2D XUV field.

Through circular RABBIT we can determine the XUV spectral phases for any two orthogonal polarization components, but both of them have additive phase constants, which need to be further characterized using the Stokes-parameter measurement. This is the key difference between the characterizations of 1D and 2D attosecond pulses.

\begin{figure}[htbp]
\centering
{\vspace{-3cm}}
\includegraphics[width=14.5cm]{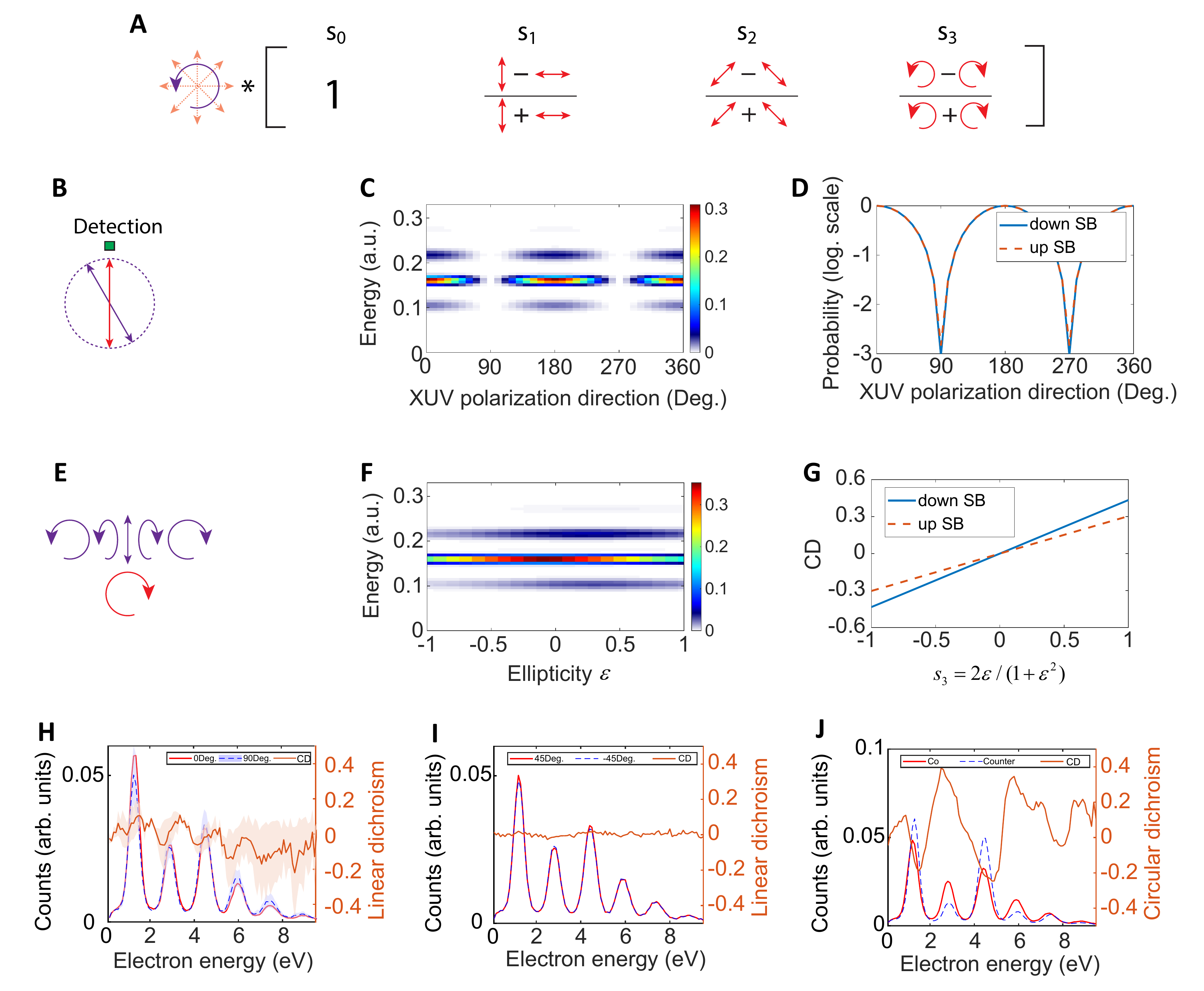}
\caption{\textbf{Electron-version Stokes-parameter measurement for characterization of the XUV polarization state, the second part of SPARROW}. \textbf{A}, The definition of the four normalized Stokes parameters. The photoionization of atoms by an XUV beam (assuming it is partially polarized) will create a replica in form of electrons in continuum states, and a synchronized IR field (fully polarized) will trigger continuum-continuum transitions of the partially polarized electrons. In the present scheme, the normalized Stokes parameters can directly be correlated to the dichroism of the sideband yield in the IR fields of orthogonal polarizations. Here, the electron continuum-continuum transitions play the same role as the optical polarizers in the conventional Stokes-parameter measurement. \textbf{B-D}, Calibration of our linear ``polarizer''. In analogy to calibrating the extinction ratio of an optical polarizer, we fix the IR polarization direction and then successively vary the polarization direction of the linearly polarized XUV field. We count only the ionization probability along the IR polarization direction. For simplicity, in our TDSE simulation we use only one order harmonic (H13). The simulated photoelectron energy spectrum as a function of the XUV polarization direction is illustrated in C and the normalized probabilities of two sidebands are illustrated in D, which shows that this method achieves an extinction ratio of 1000:1. \textbf{E-G}, Calibration of our circular ``polarizer''. The IR field is fixed at the right circular polarization state and the ellipticity of the XUV field is successively varied from -1 (counter-rotating) to +1 (co-rotating). The simulated photoelectron energy spectrum as a function of the XUV ellipticity $\epsilon$ is illustrated in f, where the left-right asymmetry is visible. This left-right asymmetry is quantified by the circular dichroism CD = $\frac{Y(\epsilon)-Y(-\epsilon)}{Y(\epsilon)+Y(-\epsilon)}$, which is shown in g as a function of the $s_3 = 2\epsilon/(1+\epsilon^2)$ parameter in this ellipticity-varying process. Based on the obtained linear mapping relation, the $s_3$ parameter can be calibrated from the measured CD. \textbf{H-J}, Measured photoelectron energy spectra and the corresponding dichroism spectra for calibration of $s_1$, $s_2$, and $s_3$ parameters, respectively. In H and I, the photoemission angle was integrated along the IR polarization direction with an opening angle of $30^\circ$, and in J the photoemission angle was integrated over $2\pi$ range. For the whole momentum spectra, see SM.}
\label{fig:figure3}
\end{figure}

We now turn to the second part of SPARROW, which is a method for the measurement of Stokes parameters that consists in detecting photoelectrons rather than photons. It is thus a convenient $in$-$situ$ measurement technique, which is not limited to specific wavelengths. As illustrated in Fig.~\ref{fig:figure3}A, the normalized Stokes parameters [$s_0$, $s_1$, $s_2$, $s_3$] of the XUV light may be correlated to sideband electron yields:
[1, $\frac{Y(0^\circ)-Y(90^\circ)}{Y(0^\circ)+Y(90^\circ)}$, $\frac{Y(45^\circ)-Y(-45^\circ)}{Y(45^\circ)+Y(-45^\circ)}$, $\frac{Y_{\mathrm{co}}-Y_{\mathrm{counter}}}{Y_{\mathrm{co}}+Y_{\mathrm{counter}}}$], where $Y(\varphi)$ is the sideband electron yield along the polarization direction at $\varphi$ angle for the linear dressing IR field, and $Y_{\mathrm{co/counter}}$ is the $\varphi$-integrated sideband electron yield when the circular dressing IR field is co/counter-rotating with respect to the XUV field. Actually, the present normalized Stokes parameters can be understood as the linear or circular dichroism (CD) which quantifies the different response to dressing IR fields of orthogonal polarizations. Compared to the standard method of Stokes parameter measurements with optical polarizers and photon detection, in our approach, the electron continuum-continuum transition serves as the XUV linear and circular ``polarizers''. If the extinction ratio of our method is sufficiently high, it is strictly equivalent to the standard method with the optical polarizer and photon detection. To calibrate the extinction ratio, we performed two groups of TDSE simulations. In the linear polarization case (Figs.~\ref{fig:figure3}B-D), we fix the IR polarization direction and then gradually vary the XUV polarization direction. The obtained sideband yield as a function of the angle spanned by the polarization vectors of two fields proves that our linear ``polarizer'' has a very high extinction ratio of 1000:1, and therefore the measured linear dichroisms can be directly used as the normalized Stokes parameters without any further calibration. In the circular polarization case (Figs. \ref{fig:figure3}E-G), the IR field is right circularly polarized and the XUV ellipticity is gradually varied from counter-rotating through linear to co-rotating. Although the sideband yield curve as a function of the XUV ellipticity is dichroic, the extinction ratio is not high enough and thus a calibration is needed. In Fig. \ref{fig:figure3}G, we show the mapping curve between the calculated circular dichroism (our $s_3$) and the standard $s_3$ parameter. Remarkably, their mapping relation is linear, allowing for a very simple calibration. In SM, we illustrate calculation results for the other sidebands which reveal the same relationship, and we also mathematically proved the linear relationship between CD and $s_3$.

\begin{figure}[htbp]
\centering
\includegraphics[width=14.5cm]{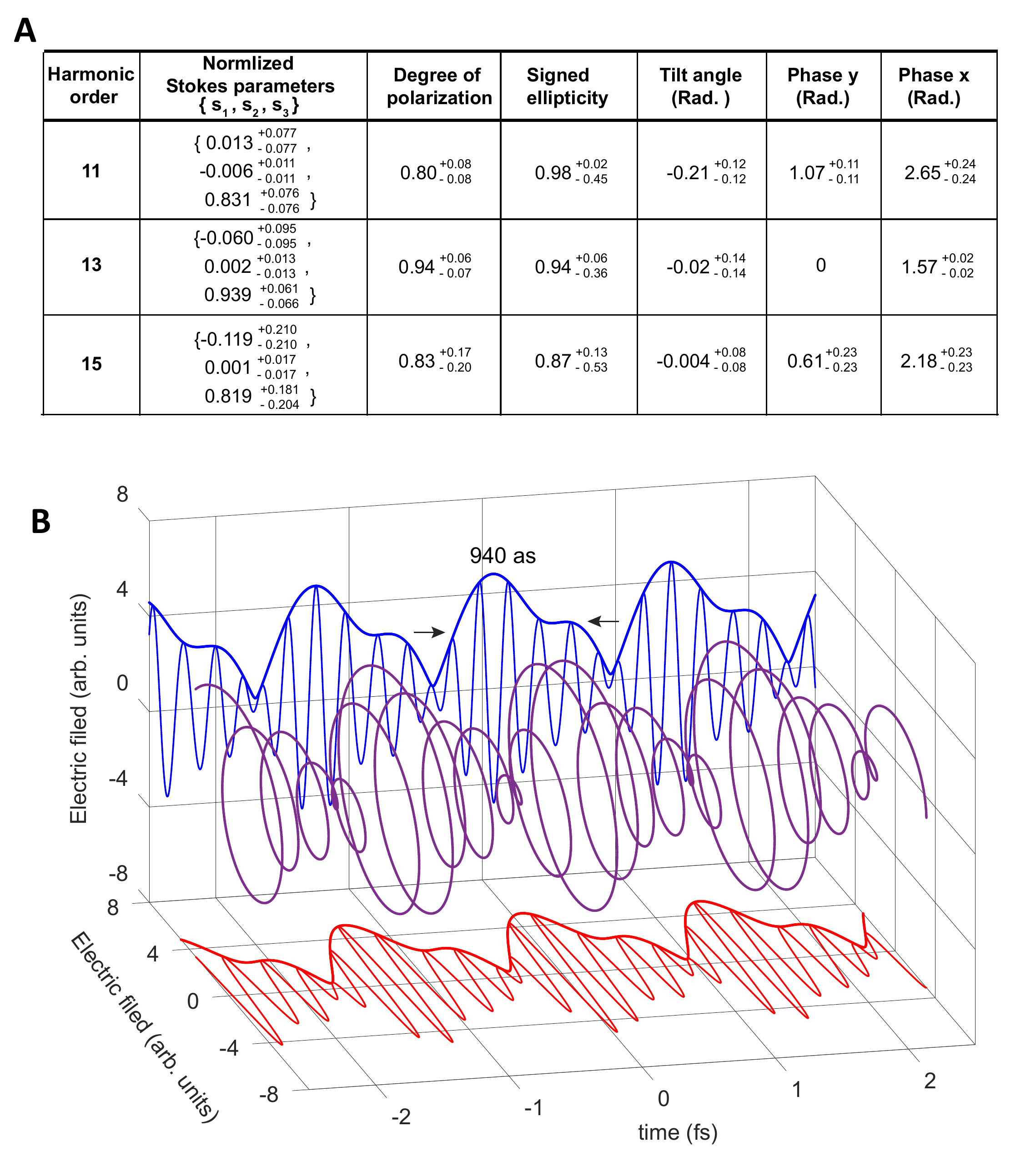}
\caption{\textbf{Pulse reconstruction of circularly polarized attosecond light field}. \textbf{A}, Table for the characterized parameters of the circular XUV field. The uncertainties of the Stokes parameters and phase y are determined by the weighted average within the corresponding sidebands, and the uncertainties of other parameters are calculated by the uncertainty propagation formula. \textbf{B}, Reconstructed attosecond temporal structure of the circular XUV field. The FWHM of the synthesised amplitude is $\sim 940$ as, which is labeled on one of its projections.  The corresponding intensity FWHM is $\sim 450$ as.}
\label{fig:figure4}
\end{figure}

Figures \ref{fig:figure3}H-J display the measured photoelectron energy spectra and the linear/circular dichroisms, corresponding to the characterization of $s_1$, $s_2$ and $s_3$, respectively. In the SM, we illustrate the whole photoelectron momentum spectrum and the corresponding dichroism spectrum. We measured that the $s_2$ parameter is very small within the range of $\pm 0.01$, and the $s_1$ parameter has a large uncertainty because of the deficiency of our electron detector along 90$^\circ$, in spite of the fact that we have corrected the energy spectrum by considering the detection efficiency. For the circular dichroism, we observed the strong dichroic signal which can reach up to $0.37$ for the lowest-order sideband. After we determined the Stokes parameters, the degree of polarization ($P_0 = \sqrt{s_1^2+s_2^2+s_3^2}$), the signed ellipticity ($\varepsilon = \mathrm{tan}(\frac{1}{2}\mathrm{asin}(\frac{s_3}{P_0}))$), and the ellipse tilt angle ($\alpha = \frac{1}{2}\mathrm{atan}(\frac{s_2}{s_1})$) can be characterized for each harmonic (see SM for details). The results are summarized in Fig.~\ref{fig:figure4}A. For each harmonic, the phase difference between two orthogonal components is determined by the tilt angle of the harmonic ellipse, i.e., $\mathrm{tan}(2\alpha) = \frac{2\varepsilon \mathrm{cos}(\varphi_y - \varphi_x)}{1 - \varepsilon^2}$, where $\varphi_y$ and $\varphi_x$ are the harmonic phases along the $y$ and $x$ directions, respectively. Therefore, in our pulse reconstruction process only one harmonic phase (here chosen to be $\varphi_y$ of H13) needs to be set arbitrarily. The $\varphi_y$ of H11 and H15 are then determined by successively adding the measured attochirp phases, and the $\varphi_x$ of three harmonics are retrieved by the ellipse tilt angles. Combining the circular RABBIT with the Stokes-parameter measurement, or performing SPARROW, a 2D XUV field can be fully characterized up to its carrier-envelope phase.

After determining these important parameters of a 2D XUV field, its corresponding temporal structure can then be retrieved accurately. Figure \ref{fig:figure4}B shows the reconstructed projections on the two directions and also the complete 2D XUV field. The pulse duration was determined to be 940~as for the full width at half maximum (FWHM) of the synthesised electric-field envelope, corresponding to a 450~as FHWM of the intensity envelope. The retrieved synthesised electric field in the time domain directly proves that the XUV pulse is purely circularly polarized from the non-collinear HHG scheme, in marked contrast with the case of three linearly polarized bursts per optical cycle in the bi-circular collinear HHG scheme \cite{Chen2016}. In the SM, we numerically demonstrate in addition that SPARROW can be applied to the characterization of an arbitrarily polarized 2D XUV field using the standard FROG-CRAB algorithm. 

In summary, we have introduced and demonstrated a straightforward plug-in device that will make circularly polarized attosecond light pulses broadly available. Our apparatus allows to accurately tune the ellipticity and select the helicity of the attosecond light pulses in an easy manner. To fully characterize the circularly polarized attosecond light pulse, we first introduced the circular RABBIT scheme to measure their spectral phases, and the electron-version Stokes-parameter measurement to characterize the polarization state, resulting in the SPARROW technique. The circular RABBIT scheme integrates the concepts of angular streaking principle with the attosecond frequency beating, enabling to perform attosecond metrology without scanning the time delay between XUV and IR fields. The electron-version Stokes-parameter approach provides a general and transparent $in$-$situ$ characterization method. Our comprehensive set of methods for attosecond metrology in circular and elliptical polarizations has the potential to extend attosecond science into the realms of chiral molecular systems, as well as magnetic and topological materials.

\clearpage
\bibliography{attobib,pop_references,newrefs}
\bibliographystyle{Science}

\section*{Acknowledgments}
We thank A. Schneider and M. Seiler for their technical support. \textbf{Funding} M.H. acknowledges the funding from the European Union’s Horizon 2020 research and innovation program under the Marie Skłodowska-Curie grant agreement No 801459 - FP-RESOMUS. This work was supported by ETH Z\"urich and the Swiss National Science Foundation through projects 200021\_172946 and the NCCR-MUST.

\noindent\textbf{Authors contributions} M.H. performed the experiments with the support of J.J.. M.H. analyzed and interpreted the data. Simulations were implemented by M.H.. H.J.W. conceived the study and supervised its realization. M.H., K.U. and H. J. W. discussed the results and wrote the paper with the input of all co-authors.

\noindent\textbf{Competing interests} None to declare. 

\noindent\textbf{Data and materials availability} All data needed to evaluate the conclusions in the paper are present in the paper or the supplementary materials.

\clearpage

\clearpage
\setcounter{page}{1}

\end{document}